\documentclass[aip,reprint]{revtex4-1}

\usepackage{amsmath}
\usepackage{graphicx}
\usepackage{upgreek}

\newcommand{\micro}{${\upmu}$}

\begin{document}

\title{Ultra-Low-Power Hybrid Light-Matter Solitons}

\author{L. Tinkler}

\author{P.M. Walker}
\email{p.m.walker@sheffield.ac.uk}

\affiliation{Department of Physics and Astronomy, University of Sheffield, S3 7RH, Sheffield, United Kingdom}

\author{D.V. Skryabin}
\affiliation{Department of Physics, University of Bath, BA2 7AY, Bath, United Kingdom}

\author{A. Yulin}
\affiliation{Centro de Fisica Teorica e Computacional, Universidade de Lisboa, P-1649003 Lisbon, Portugal;
ITMO University 197101, Kronverksky pr. 49, St. Petersburg, Russian Federation}

\author{B. Royall}

\affiliation{Department of Physics and Astronomy, University of Sheffield, S3 7RH, Sheffield, United Kingdom}

\author{I. Farrer}
\author{D.A. Ritchie}

\affiliation{Cavendish Laboratory, University of Cambridge, CB3 0HE, Cambridge, United Kingdom}

\author{D.N. Krizhanovskii}
\author{M.S. Skolnick}

\affiliation{Department of Physics and Astronomy, University of Sheffield, S3 7RH, Sheffield, United Kingdom}

\begin{abstract}
New functionalities in nonlinear optics will require systems with giant optical nonlinearity as well as compatibility with photonic circuit fabrication techniques. Here we introduce a new platform based on strong light-matter coupling between waveguide photons and quantum-well excitons. On a sub-millimeter length scale we generate sub-picosecond bright temporal solitons at a pulse energy of only 0.5 pico-Joules. From this we deduce an unprecedented nonlinear refractive index 3 orders of magnitude larger than in any other ultrafast system. We study both temporal and spatio-temporal nonlinear effects and for the first time observe dark-bright spatio-temporal solitons. Theoretical modelling of soliton formation in the strongly coupled system confirms the experimental observations. These results show the promise of our system as a high speed, low power, integrated platform for physics and devices based on strong interactions between photons.
\end{abstract}

\maketitle 
The introduction of strong nonlinearity into photonic circuits has many applications ranging from optical information processing~\cite{polariton_transistor,Amo_spin_switch,neuron} and soliton physics~\cite{kivshar,Lumer_Topological_Solitons} to the study of inter-particle interactions~\cite{Manela2010_nonlinear_hofstadter_butterflies,Hafezi_Topological_Physics_With_Light,Umucalilar2012_Fractional_Quantum_Hall,Lumer_Topological_Solitons,Carusotto2009_Fermionized_Photon_Cavity_Array,Tercas2014,Carusotto_RevModPhys2013} in photonic analogs of important physical systems such as photonic topological insulators\cite{Umucalilar2011,Hafezi2011_robust_optical_delay_lines_with_topological_protection,Fang_Mag_Field_Dynamic_Modulation,Hafezi2013_Imaging_topological_edge_states_in_silicon_photonics}, optical analogs of quantum hall systems. In this work we address the problem of introducing strong optical nonlinearity by studying ultra-low-power optical solitons in a novel slab waveguide geometry we recently introduced~\cite{Walker2012} where photons are strongly coupled to quantum-well excitons. Temporal optical solitons are one of the most fundamental nonlinear phenomena in optics. They arise due to a competition between group velocity dispersion (GVD), where different frequency components accumulate different phase during propagation, and nonlinear phase modulation where intensity dependent phase shifts can exactly balance those due to GVD~\cite{kivshar}. They have been proposed for applications~\cite{kivshar} and observed~\cite{Mollenauer_Fibre_Solitons} in long-haul communications lines and more recently investigated in the micro- and nano-scale waveguides suitable for design of all optical processing chips \cite{Zhang2007,skr1,Colman2010,bragg,Blanco-Redondo2014}. On-chip applications require that solitons form and interact over small distances, which requires that both the GVD and nonlinear effects develop on very short length scales. It is also important for information processing applications that the underlying nonlinearity responds on picosecond or faster timescales. Achieving short nonlinear length scales while minimising power requirments necessitates very large nonlinearity. To generate significant GVD, photonic crystal band engineering and semiconductor photonic wires have been used in a number of schemes\cite{Zhang2007,skr1,Colman2010,bragg,Blanco-Redondo2014}.

In this work we simultaneously achieve giant nonlinearity and GVD by exploiting the unique optical properties of exciton-polaritons~\cite{book}. These quasi-particles are the quantum eigenmodes formed when photons couple strongly to a quantum-well exciton resonance. They have characteristics coming from both their light and matter constituents, with the proportion depending on the detuning from resonance. The photonic component contributes long lifetime and detailed control over system properties via photonic circuit fabrication techniques~\cite{Wertz2012,Jacqmin2014}. Matter-like interparticle scattering leads to enormous nonlinearity which responds on an ultrafast timescale. We demonstrate formation of bright sub-picosecond temporal polariton solitons with pulse energies of less than 0.5 pJ. From the soliton formation threshold we deduce a nonlinear refractive index three orders of magnitude larger than in any other ultrafast optical system, which verifies the highly nonlinear nature of our platform. We show that the frequency dependent transition of polaritons from photon-like to matter-like leads to giant group velocity dispersion on the order of 1000 ps$^2$ m$^{-1}$. This is comparable to that achieved in fiber Bragg gratings and photonic crystal waveguides\cite{bragg,Colman2010,Blanco-Redondo2014}. As we achieve the GVD necessary for soliton formation without the need for transverse photonic confinement we are able to study spatio-temporal effects as well as purely temporal ones. We experimentally demonstrate dark-bright spatio-temporal solitons, which to our knowledge have not previously been seen in any experimental setting with either photons or polaritons. These are solitons where the intensity is peaked in the propagation direction and extended in the transverse direction apart from a dark notch. Note that bright\cite{Sich2011,Tanese2013} and spatial dark\cite{Amo2011} exciton-polariton solitons have previously been reported in strongly coupled Bragg microcavities. Compared to these our system has the advantages of an order of magnitude faster soliton velocity, simplicity of design and fabrication, compatibility with existing designs of high performance waveguide photonic circuit elements and, crucially, does not depend on external continuous wave pumping to balance loss for soliton formation to be possible. Apart from these, the major advantages of our platform for nonlinear optics are the combination of very large nonlinearity and ultra-fast temporal response. The intrinsic nature of the large dispersion is also advantageous for on-chip soliton formation, pulse compression~\cite{Blanco-Redondo2014} and the study of two-dimensional effects.

\begin{figure}
\centering
\includegraphics[width=8.5 cm]{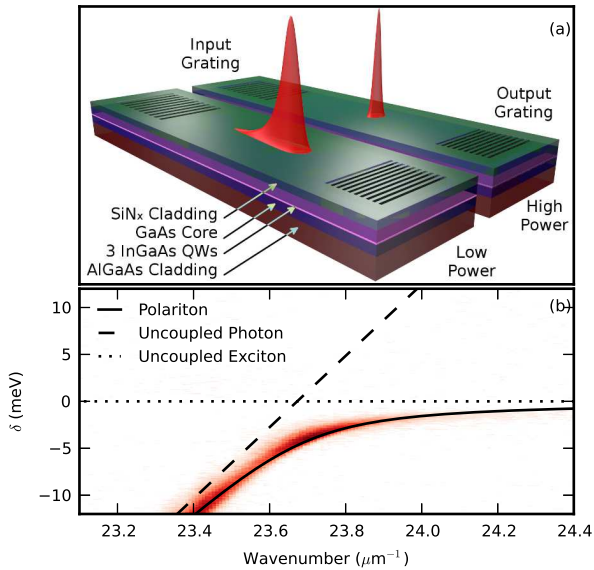}
\caption{(a) Schematic diagram of waveguide (the same waveguide is shown with pulses under low power and high power excitation conditions). (b) Angle resolved photoluminescence spectrum showing emission from the lower polariton branch in red. The fitted polariton dispersion (solid line) is shown with the uncoupled exciton and photon modes which would exist for zero light-matter coupling (dotted and dashed lines). The horizontal and vertical axes are respectively the wavenumber of the guided mode and the detuning $\delta$ between polariton and exciton energies.}
\label{fig1}
\end{figure}

The structures we use in this work are planar AlGaAs waveguides with quantum wells parallel to the substrate~\cite{Walker2012}, see schematic diagram in Fig.~\ref{fig1}a). Diffraction in the crystal growth direction is prevented by total internal reflection. As the photonic losses are moderate and broadening of the exciton line is suppressed (the temperature is 10K) the system is in the strong coupling regime and exciton-polaritons are formed. The sample is excited by pulses with a square spectral envelope of 5.5meV corresponding to an approximately 350 femto-second (fs) temporal full width at half maximum (FWHM). The pulses are coupled in at near-zero incidence angle through the input grating, propagate 600 \micro m and couple out through the output grating. Fig.~\ref{fig1}b) shows the polariton dispersion relation determined by angle-resolved photoluminescence spectroscopy\cite{Walker2012}. As the momentum increases along the lower polariton branch the polaritons become more matter like and their group velocity approaches zero.

In order to model the experimental results we describe a light pulse interaction with a narrow exciton resonance using the Maxwell-Lorentz system. Any group velocity dispersion (GVD) related to the waveguide geometry is negligible over the spectral range close to the exciton resonance and in comparison with the GVD induced by the photon exciton coupling. The resulting set of equations accounting for the dominant nonlinearity originating in the two-body exciton-exciton interaction is
\begin{subequations} \label{eq:max_lorentz}
\begin{align}
2i\beta_e(\partial_z+v_g^{-1}\partial_t+\gamma_{p})A+\partial^2_xA=-k_e^2\psi, \label{eq:max_lorentz_a} \\
-2i(\partial_t+\gamma_{e})\psi=\kappa A-g |\psi|^2\psi. \label{eq:max_lorentz_b}
\end{align}
\end{subequations}
Here $\beta_e$ and $v_g$ are the propagation constant and the group velocity of the photonic waveguide mode at the exciton frequency $\omega_e$ and $k_e=\omega_{e}/c$. $z$ and $x$ are the coordinates, respectively, along and across the propagation direction. The waveguide width along $x$ is much larger than the diffractive spread achieved over its length. $t$ is time, $A$ is the amplitude of the photonic mode, which is tightly bound in the direction perpendicular to the waveguide plane, and $\psi$ is the excitonic polarization scaled to be in the same units as $A$. $\gamma_{p,e}$ are the photon and the exciton coherence decay rates and $\kappa$ is the rate of the light matter coupling. For plane-waves $A$ and $\psi$ proportional to $e^{iQz-i\delta t}$, where $Q$ is a propagation constant offset from $\beta_e$ and $\delta$ is a frequency offset from $\omega_e$, the lossless dispersion law is $Q=v_g^{-1}(\delta-\Omega_R^2/\delta)$. Here $2\Omega_R=k_e\sqrt{\kappa v_g/\beta_e}=9$meV is the energy splitting between the two polariton branches at $Q=0$. The GVD parameter $\beta_2=\partial_{\delta}^2Q=-2\Omega_R^2/(\delta^3v_g)$. Thus the GVD is normal (i.e. $\beta_2>0$) for pulses with central frequency $\delta_c$ detuned down from the resonance ($\delta_c<0$) towards the lower polariton branch and anomalous ($\beta_2<0$) otherwise. In our experiment the pulse central (carrier) frequency $\hbar\delta_c$ varies from $-7.6$meV to $-10.4$meV, so that we operate on the part of the lower polariton branch where the polaritons are between 74\% and 84\% photonic and $\beta_2$ is between 400 and 1000 ps$^2$ m$^{-1}$ (see Fig.~\ref{fig1}b).

Within our experimental and theoretical frameworks we study both quasi-one dimensional effects when diffraction along $x$ is not important and two-dimensional effects, when diffraction is relevant for the physics of the process. Mathematically, the system under consideration is similar to the equations used to study gap solitons \cite{kivshar} and solitons with dispersion provided by material resonances of different physical origin \cite{gabit,maim}. Note that the latter type of solitons has mostly been studied theoretically and the experimental studies have been focused on so-called self-induced transparency solitons in atomic gases \cite{sitexp}. Spatio-temporal solitons are a separate large area of research, where the most convincing experimental measurements of quasi-solitonic effects have been demonstrated in $\chi^{(2)}$ crystals with very challenging dispersion control \cite{chi2} and in arrays of coupled waveguides \cite{jena}. Previous efforts to use nanophotonic systems to observe spatio-temporal effects \cite{bath} have not been very successful.

\begin{figure*}
\centering
\includegraphics[width=16.9 cm]{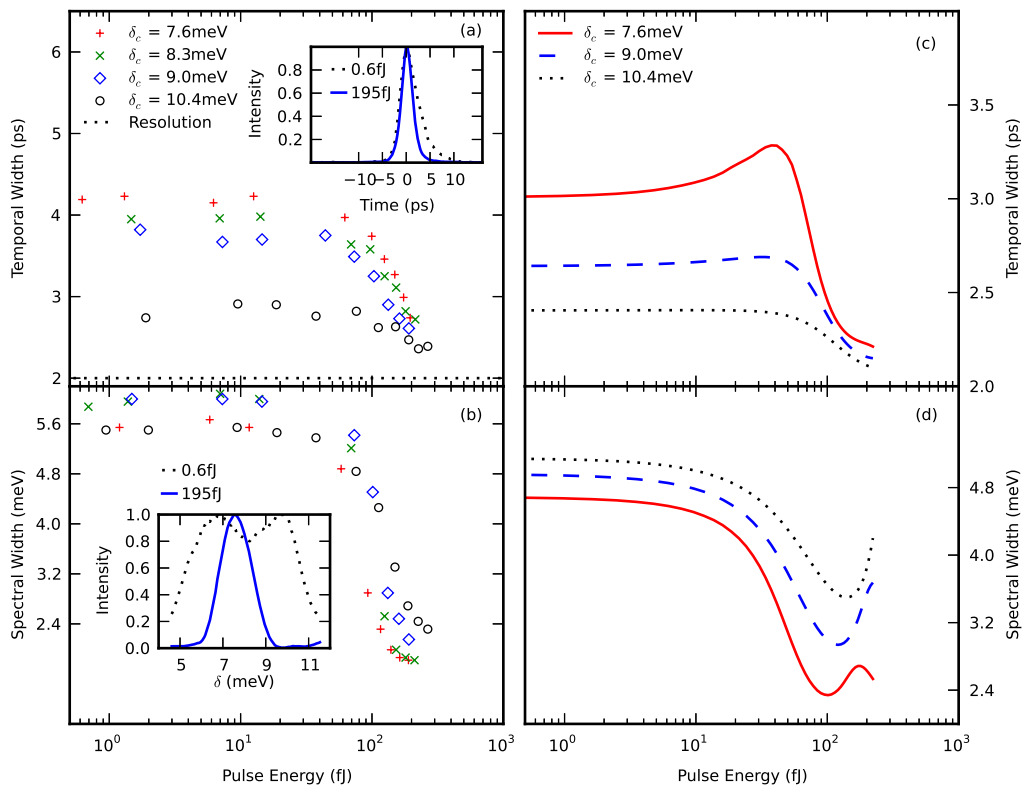}
\caption{Comparison of experimentally measured and numerically simulated temporal and spectral pulse widths (FWHM) at the output grating. (a) Experimental pulse temporal duration vs. pulse energy inside the waveguide for a number of detunings $\delta_c$ of the central pulse frequency from the exciton. Inset temporal profiles at high and low pulse energy. (b) Experimental pulse spectral width vs. pulse energy for the same detunings. Inset spectral profiles at high and low pulse energy. (c,d) Numerically simulated temporal and spectral widths vs. pulse energy for similar excitation conditions as in the experiment. The semi-quantiative agreement indicates that solitons are observed in the experiment.}
\label{fig_temporal_results}
\end{figure*}

We proceed by describing our results on observation of one dimensional temporal solitons. We first present experimental measurements which demonstrate temporal soliton formation and then confirm this by comparing the data with numerical modelling. Dependencies of the output pulse width vs total output pulse energy are shown in Fig.~\ref{fig_temporal_results}a) for several detunings of the central pulse frequency from the exciton, $\delta_c$. The detector resolution causes the incident 350fs pulse to be measured as 2.0ps. Below a threshold pulse energy the measured FWHM after propagation though the waveguide is between 2.7 to 4.2 ps depending on $\delta_c$, which indicates significant temporal speading due to the group velocity dispersion. The inset of Fig.~\ref{fig_temporal_results}a) shows the temporal profile of the pulse collected from the output grating for pulse energies above and below the threshold. One should note that the pulse shape is highly asymmetric below threshold and becomes symmetric above threshold. Fig.~\ref{fig_temporal_results}b) shows the corresponding output spectral widths while the inset shows pulse spectra above and below threshold. For low pulse energies the output spectrum reflects that of the incident pulse slightly modified by the frequency dependent coupling efficiency of the grating coupler (see inset). As the pulse energy is increased above threshold the pulse becomes significantly shorter and more symmetrical in time until, for the strongest pulses, the measured FWHM reaches a minimum between 2.4ps and 2.7ps, depending on $\delta$. At the same time the spectrum shows strong narrowing about the central frequency reaching values in the range 1.8 to 2.3meV, as shown in Fig.~\ref{fig_temporal_results}b). During this process the total power collected from the output grating remains linear with pump power to within a few percent indicating that the spectral and temporal compression are due to a coherent transfer of energy between frequency components rather than a nonlinear absorption or filtering effect. Such coherent spectral and temporal compressions occuring at a critical input energy and levelling off with further energy increase are characteristic of the transition from dispersive propagation to soliton formation.

\begin{figure}
\centering
\includegraphics[width=8.5 cm]{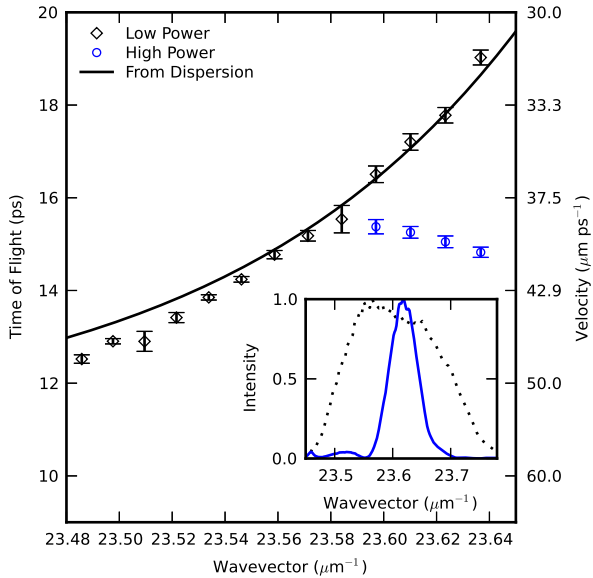}
\caption{Measured time of flight of different wavevector components of pulse at low power (black diamonds)
and high power (blue circles) and time of flight extracted from curvature of polariton dispersion (solid curve).
Inset shows pulse intensity as a function of wavevector at high (solid blue line) and low (dotted black line) power.}
\label{fig3}
\end{figure}

Another key signature of soliton formation is linearisation of the soliton dispersion relation resulting in a single group velocity for all soliton components. Figure \ref{fig3} shows the time of flight of different wavevector components of the pulse above and below threshold. Below threshold the time of flight is simply determined by the polariton dispersion relation with polaritons close to resonance travelling slower, see Fig.~\ref{fig1}(b). Above threshold, however, there is a very clear departure from this trend. To within the experimental error the time of flight becomes constant with wavevector across the range where there is significant intensity. This should be compared with a difference of 4ps over the same wavevector range below threshold. The dispersion is therefore cancelled out above threshold confirming the formation of a soliton. The soliton velocity of 39 \micro m ps$^{-1}$ is faster than that of linearly propagating polaritons at the same energies but significantly slower than the 58 \micro m ps$^{-1}$ expected for uncoupled photons. This confirms that the strong light-matter coupling is preserved during the propagation of the high power pulse. To summarise, the experimental evidence for soliton formation consists of strong coherent temporal and spectral narrowing of the output pulses at a critical energy and a strong reduction in the spread of velocities among different wavenumber components of the pulses.

Soliton formation requires the characteristic length scale $L_{NL}=T/(\gamma E)$ for the buildup of nonlinear phase to be comparable to the dispersion length. Here, $T=$350fs is the pulse width, $E$ is the characteristic pulse energy for the soliton formation and $\gamma$ is the nonlinear waveguide parameter \cite{kivshar}. The dispersion length is given by $L_D = T_0^2/\left|\beta_2\right|\simeq 44-110\mu$m where $T_0 \simeq T/1.665$ is the $1/e$ pulse half width. We note that solitons in any lossy system will lose intensity as they propagate with a characteristic loss length $L_{loss}\simeq$400\micro m in our system. Provided that $L_{NL}=L_D<L_{loss}$ the nonlinearity will adiabatically adjust the pulse width to compensate the loss and solitons may still be expected. From these considerations we now estimate the effective nonlinear refractive index in our system. Taking the threshold pulse energy at the waveguide exit as 100fJ and $L_{loss}=$ 400\micro m we estimate $E=450$fJ at the input. Setting $L_{NL}=L_D$ we obtain $\gamma=T/(E L_D)\sim$ -18,000/(Wm). This may be related to the nonlinear refractive index by $\gamma = n_2 k_0 / A_{eff}$ where $k_0$ is the vacuum wavenumber and $A_{eff} = $ 6.6 \micro m$^2$ is the effective mode area~\cite{Agrawal_Nonlinear_Fibre_Optics}. We therefore obtain $n_2 = $-1.6x10$^{-14}$ m$^2$W$^{-1}$. This is more than three orders of magnitude larger than 6x10$^{-18}$ m$^2$W$^{-1}$ in silicon~\cite{Blanco-Redondo2014,Zhang2007}and InGaP~\cite{Colman2010} which have recently been used in a suspended membrane photonic crystal geometry in two of the most promising demonstrations of solitons for integrated optics until now. In those systems solitons are formed at pulse energies of 12pJ for InGaP~\cite{Colman2010} and 9pJ for silicon~\cite{Blanco-Redondo2014} which are 1.3-1.4 orders of magnitude higher than in our system. We note that in those works the effective mode areas are smaller ($\sim$0.5\micro m$^2$) and the dispersion lengths are much longer (3.6mm and 1mm) than in our system both of which can be used to reduce the threshold, the latter at the expense of increased circuit size.

\begin{figure}
\centering
\includegraphics[width=8.5 cm]{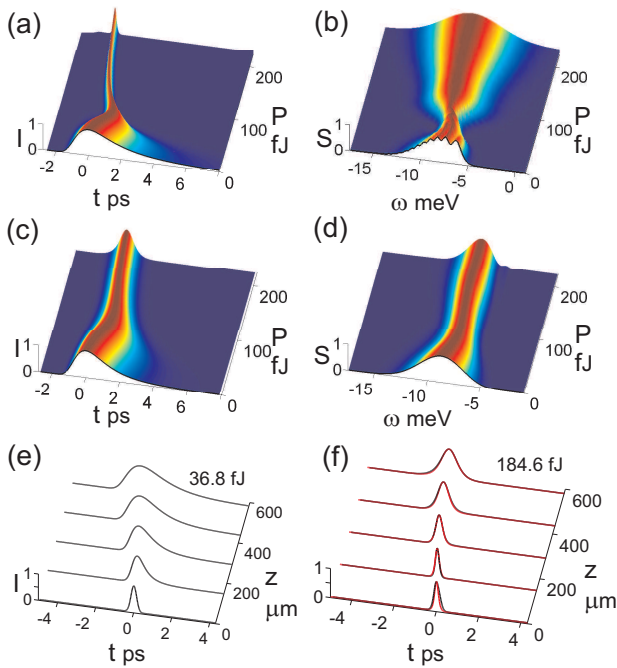}
\caption{Numerically computed time and frequency domain pulse profiles at the end of the waveguide for the lossless (a,b) and lossy (c,d) cases. Evolution of the pulse along the length of the waveguide incuding loss for pulse energies below (e) and above (f) the soliton formation theshold respectively. In (f) the red curves are analytic soliton solutions.}
\label{fig:theory}
\end{figure}

Figs.~\ref{fig:theory}(a,b) show numerically computed time domain and spectral pulse profiles vs pulse energy at the end of the waveguide under the assumption that loss is neglected. By tracing the evolution of the pulse along the waveguide we have confirmed that exact temporal solitons are excited for pulse energies above $\simeq 70$fJ. Experimental values of the losses introduced into the modeling modify the pulse dynamics only quantitatively, see Figs.~\ref{fig:theory}(c,d). They accelerate pulse broadening and shift the threshold for formation of solitons towards higher energies, $\simeq 120$fJ. Figs.~\ref{fig:theory}(e,f) show evolution of the pulse profile along the waveguide with the real losses below (e) and above (f) the soliton formation threshold. For every profile shown in (f) the red curves indicate a matching analytical soliton solution, see Methods, thus confirming that we are dealing with adiabatically decaying solitons. For comparison with experimental results, the modelled dependencies of the temporal and spectral widths of the output pulses on pulse energy are shown in Figs.~\ref{fig_temporal_results}(c,d). The 2ps resolution of the streak camera has been included in the temporal widths. As in the experimental case there is a temporal compression of up to 1ps depending on detuning which occurs over a few tens of femto-Joules pulse energy. Also as in the experiment the spectral widths reduce by $\sim$ 2meV over the same pulse energy range. This good semi-quantitative agreement between experiment and theoretical soliton solutions futher confirms the observation of temporal solitons.

We now shift our attention to the conditions where diffraction in the transverse direction is important. Since the polariton-polariton interaction is repulsive, corresponding to defocussing nonlinearity, a gaussian beam is expected to exhibit enhanced diffraction as the power increases. As well as this defocussing the system is also expected to support dark spatial solitons \cite{Chen2012,Kivshar1998}. Ideally, these consist of a high intensity background extending to infinity in the direction transverse to propagation with a dark notch at the centre whose width is defined by the background intensity. In analogy to temporal solitons the fundamental spatial soliton is formed when the nonlinear length matches the diffraction length so that diffraction is balanced by self phase modulation. The intrinsically large GVD in our system makes the dispersion and diffraction lengths comparable so that both may be compensated at the same low pulse energy and thus form spatio-temporal solitons. Before proceeding it is worth mentioning that if transverse effects are undesirable for a particular application they may be removed simply by employing transverse confinement such as a waveguide ridge. Staying in the planar geometry, the excitation beam transverse profile was modified to excite a dark soliton following the approach used in Ref.~\onlinecite{Shandarov2000}. The initially gaussian beam was passed through a thin-film phase plate in order to introduce a ${\pi}$ phase jump at its centre. The intensity distribution at the input grating then consisted of two bright lobes with a dip in intensity between them where the phase is discontinuous.

\begin{figure}
\centering
\includegraphics[width=8.5 cm]{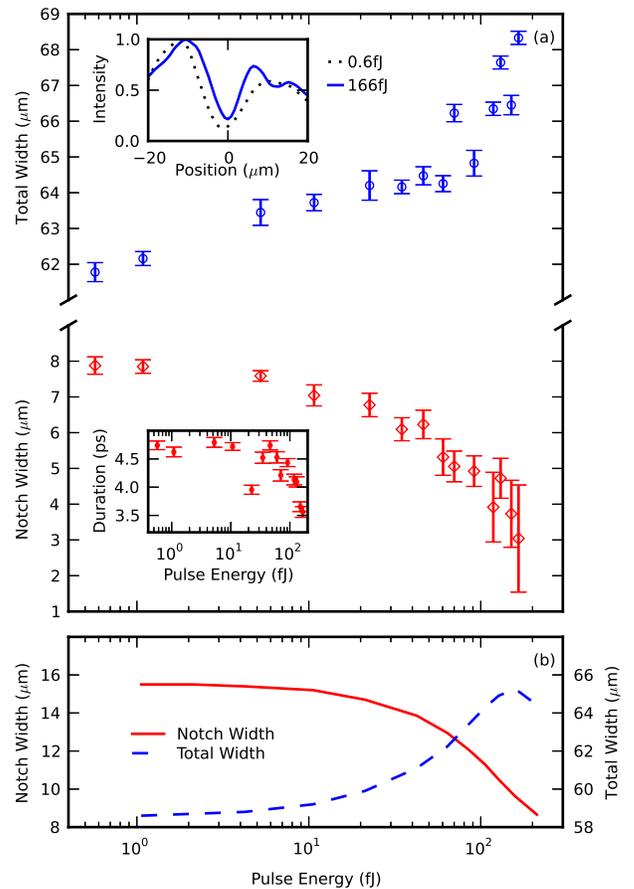}
\caption{Experimentally measured and numerically calculated transverse widths of the total distribution and of the dark notch at the output grating as a function of pulse energy. (a) Experimental widths of the dark notch (bottom) and the total distribution (top). Inset at the top is the measured transverse profile at high and low pulse energy. Inset at the bottom is the pulse temporal duration vs. pulse energy. (b) Numerically simulated widths of the transverse profile for similar excitation conditions as in the experiment.}
\label{fig:bright_dark}
\end{figure}

Figure \ref{fig:bright_dark} shows the width of the full distribution and of the dark notch measured at the output grating as a function of pulse energy. The upper inset of the figure shows the transverse profile for high and low powers. As the power is increased the total width of the distribution increases by 6 microns due to nonlinear defocussing as expected. In contrast, the width of the intensity notch decreases by 5 microns. A simple defocussing effect would be expected to enhance the diffraction of the notch as well as the sidelobes. The fact that the notch narrows indicates that we form a dark soliton. In fact, the combination of sidelobe broadening and notch narrowing represents the distribution gradually approaching that of the ideal dark soliton as the nonlinear interaction becomes important. At the same powers as the spatial soliton forms the temporal duration decreases from 4.7ps to 3.6ps. Thus we have observed clear experimental signatures of spatio-temporal dark-bright solitons. These solitons have been previously studied only theoretically\cite{prest,canada}. We have also numerically modeled formation of these solitons and found our results to be in very good agreement with the experiment. The numerically computed transverse widths are shown in Fig.~\ref{fig:bright_dark}b). As in the experimental case the nonlinear interaction widens the total distribution and narrows the dark notch by several microns with the rate of change of width increasing with power. The good semi-quantiative agreement of these trends again confirms that the experiment observes spatio-temporal solitons.

In conclusion, we have demonstrated formation of picosecond light-matter solitons with unprecedented low pulse energy density. We have studied both temporal and spatio-temporal nonlinear effects and observed the dark-bright spatio-temporal soliton for the first time. The solitons exist in a novel regime where intrinsically large dispersion is balanced by a nonlinearity several orders of magnitude larger than in any other ultrafast system. These properties allow nonlinear phases to accumulate and be controlled over short distances consistent with on-chip applications at low power. The waveguide exciton-polariton system therefore represents an important platform for high speed integrated optics requiring large nonlinearity.

\section{Acknowledgements}
We acknowledge support from EPSRC Programme Grant EP/J007544/1, ERC Advanced Grant Excipol, 320570 and the Leverhulme Trust.

\section{Methods}
\subsection{Experiment}
Our system comprises a planar waveguide with GaAs core, AlGaAs bottom cladding and silicon nitride top cladding. Three shallow In$_{\textnormal{0.04}}$Ga$_{\textnormal{0.96}}$As quantum wells are embedded in the core parallel to waveguide plane and at the peak of the electric field density. Details of the structure are the same as those in Ref.~\onlinecite{Walker2012} except for the use of three quantum wells. The transverse-electric (TE) polarised guided optical mode couples strongly to the quantum well 1s heavy-hole exciton at wavelength 839.6nm to form TE polarised polaritons. The vacuum Rabi splitting is 9meV. The linear loss length of 400 \micro m is due to a combination of photon tunneling through the cladding and optical absorption by the GaAs core material and tail of the exciton line. Excitation laser pulses were generated by spectrally filtering 100fs pulses from a mode-locked titanium doped sapphire laser with a diffractive pulse shaper employing an adjustable slit. The laser repetition rate was 80MHz. The pulses were injected into the waveguide through the grating couplers of period 250nm by tuning their incidence angle and central energy to correspond to those of the guided mode. Coupling was maximised at around 10 degrees by monitoring the emission from the output grating. Light coming from the sample was collected by an objective lens and spatially filtered so that only light from the output grating was detected. The temporal profile of the pulses was measured by projecting an image of the output grating onto the entrance slit of a streak camera. The streak camera has a temporal resolution of 2.0ps, measured using the reflection of the excitation laser pulse from an unpatterned part of the sample surface. The spectrally resolved emission was measured by projecting the fourier plane of the objective onto the entrance slit of an imaging spectrometer and integrating the light with a CCD detector. The pulse energy was determined by focussing all light from the output grating onto a commercial photodiode-based power meter and dividing the average beam power by the laser repetition rate. A ratio of 1:4 for the power scattered towards the power meter and that lost in the substrate was determined by modelling pulse outcoupling through the grating using a two-dimensional finite-difference-time-domain (FDTD) method and has been included to obtain the total pulse energy before outcoupling. The wavevector resolved time-of-flight measurements were performed by projecting the fourier plane of the objective onto the entrance slit of the streak camera. The final lens was scanned in order to align each wavevector component with the slit. The photoluminescence emission presented in Fig~\ref{fig1}b) was measured using continuous wave laser excitation at 685nm wavelength.

\subsection{Analytical soliton solution}
One dimensional  temporal  solitons can be found analytically. Neglecting losses, $\gamma_p=\gamma_e=0$, and disregarding  diffraction $\partial_x^2 A=0$, the equations \ref{eq:max_lorentz} become
\begin{subequations} \label{eq:max_lorentz_temp}
\begin{align}
2i\beta_e(\partial_z+v_g^{-1}\partial_t)A=-k_e^2\psi, \label{eq:max_lorentz_a_temp} \\
-2i\partial_t\psi=\kappa A-g |\psi|^2\psi. \label{eq:max_lorentz_b_temp}
\end{align}
\end{subequations}

The conservative soliton solutions are parameterized by  the soliton frequencies $\delta_s$ and velocities $v_s$ and can be sought in the form
\begin{equation} \label{eq:analitycal_solution}
\psi=\rho_{\psi}\cdot e^{i\delta_s \tau + i\theta_{\psi}}, \quad A=\rho_{A}\cdot e^{i\delta_s \tau + i\theta_{A}},
\end{equation}
where $\rho_{\psi}(\tau)$, $\rho_{A}(\tau)$, $\theta_{\psi}(\tau)$ and $\theta_{A}(\tau)$ are real functions of $\tau=t-z/v_s$.
Substituting the anzats (\ref{eq:analitycal_solution}) into equation (\ref{eq:max_lorentz_temp}) we obtain the expressions for amplitudes $\rho_{\psi}$, $\rho_A$ of the field. They read
\begin{subequations} \label{eq:analitycal_solution_ampl}
\begin{align}
\rho_{\psi}^2=\frac{1}{g \mu}\frac{4\sigma(1-\sigma^2\mu^2\delta_s^2)}{\cosh(\frac{\sigma}{\mu}\sqrt{1-\sigma^2\mu^2\delta_s^2} \tau)+\sigma\mu\delta_s} , \label{eq:analitycal_solution_ampl_psi} \\
\rho_A=\sigma \rho_{\psi}, \label{eq:analitycal_solution_ampl_A}
\end{align}
\end{subequations}
where
$\sigma^2=\frac{v_s}{v_s-v_g}$ and $\mu^2=\frac{\beta_e}{\kappa v_g k_e^2}$.

The expressions for the phases $\theta_{\psi}$ and $\theta_A$ of the fields are given by the integrals
\begin{subequations} \label{eq:analitycal_solution_phase}
\begin{align}
\theta_{\psi}=  \frac{1}{2}\sigma^2 \delta_s\tau-\frac{3 g}{8}\int_{-\infty}^{\tau}\rho_{\psi}^2 d\tau^{\prime}, \label{eq:analitycal_solution_phase_psi} \\
\theta_{A}= \frac{1}{2}\sigma^2 \delta_s\tau-\frac{g}{8}\int_{-\infty}^{\tau}\rho_{A}^2 d\tau^{\prime}. \label{eq:analitycal_solution_phase_A}
\end{align}
\end{subequations}

\end{document}